\documentclass[12pt,leqno]{article}
\usepackage{amsmath, amsthm, amsfonts, amssymb, color}
\usepackage{mathrsfs}
\theoremstyle{definition}

\newcommand{\scr}[1]{\mathscr #1}
\definecolor{wco}{rgb}{0.5,0.2,0.3}

\numberwithin{equation}{section} \theoremstyle{remark}

\newcommand{\ua}{\uparrow}

\title{{\bf Heat Kernel   for Fractional Diffusion Operators with Perturbations}\footnote{Supported in
 part by  Lab. Math. Com. Sys., NNSFC(11131003), SRFDP and the Fundamental Research Funds for the Central Universities.}
}
\author{
{\bf Feng-Yu Wang$^{a),b)}$ and Xi-Cheng Zhang$^{c)}$ }\\
\footnotesize{a) School of Mathematical Sciences,
Beijing Normal University, Beijing 100875, China}\\
 \footnotesize{b) Department of Mathematics,
Swansea University, Singleton Park, SA2 8PP, UK}\\
\footnotesize{c) School of Mathematics and Statistics,
Wuhan University, Wuhan 430072, China}}
\begin{document}
\def\R{\mathbb R}  \def\ff{\frac} \def\ss{\sqrt} \def\B{\mathbf
B}
\def\N{\mathbb N} \def\kk{\kappa} \def\m{{\bf m}}
\def\dd{\delta} \def\DD{\Delta} \def\vv{\varepsilon} \def\rr{\rho}
\def\<{\langle} \def\>{\rangle} \def\GG{\Gamma} \def\gg{\gamma}
  \def\nn{\nabla} \def\pp{\partial} \def\EE{\scr E}
\def\d{\text{\rm{d}}} \def\bb{\beta} \def\aa{\alpha} \def\D{\scr D}
  \def\si{\sigma} \def\ess{\text{\rm{ess}}}
\def\beg{\begin} \def\beq{\begin{equation}}  \def\F{\scr F}
\def\Ric{\text{\rm{Ric}}} \def\Hess{\text{\rm{Hess}}}
\def\e{\text{\rm{e}}} \def\ua{\underline a} \def\OO{\Omega}  \def\oo{\omega}
 \def\tt{\tilde} \def\Ric{\text{\rm{Ric}}}
\def\cut{\text{\rm{cut}}} \def\P{\mathbb P} \def\ifn{I_n(f^{\bigotimes n})}
\def\C{\scr C}      \def\aaa{\mathbf{r}}     \def\r{r}
\def\gap{\text{\rm{gap}}} \def\prr{\pi_{{\bf m},\varrho}}  \def\r{\mathbf r}
\def\Z{\mathbb Z} \def\vrr{\varrho} \def\ll{\lambda}
\def\L{\scr L}\def\Tt{\tt} \def\TT{\tt}\def\II{\mathbb I}
\def\i{{\rm in}}\def\Sect{{\rm Sect}}\def\E{\mathbb E} \def\H{\mathbb H}
\def\M{\scr M}\def\Q{\mathbb Q} \def\texto{\text{o}} \def\LL{\Lambda}\def\K{\scr K} \def\KK{\mathbb K}
\def\Rank{{\rm Rank}} \def\B{\scr B} \def\i{{\rm i}} \def\HR{\hat{\R}^d}
\def\parallel{//} \def\div{{\rm div}}

\maketitle
\begin{abstract} Let $L$ be an elliptic  differential operator on a complete connected Riemannian manifold $M$ such that the associated  heat kernel has two-sided Gaussian
bounds as well as a Gaussian type gradient estimate. Let $L^{(\aa)}$
be the $\aa$-stable subordination of $L$ for $\aa\in (1,2).$ We
found some classes $\mathbb K_\aa^{\gg,\bb} (\bb,\gg\in [0,\aa))$ of
time-space functions  containing the Kato class, such that for any
measurable $b: [0,\infty)\times M\to TM$ and $c: [0,\infty)\times
M\to M$ with $|b|, c\in \mathbb K_\aa^{1,1},$ the operator
$$L_{b,c}^{(\aa)}(t,x):= L^{(\aa)}(x)+ \<b(t,x),\nn \cdot\> +c(t,x),\ \ (t,x)\in [0,\infty)\times M$$ has a unique heat kernel $p_{b,c}^{(\aa)}(t,x;s,y), 0\le s<t, x,y\in M$, which is jointly continuous and satisfies
\beg{equation*}\beg{split} &\ff{t-s}{C\{\rr(x,y)\lor (t-s)^{\frac{1}{\aa}}\}^{d+\aa}}\le p_{b,c}^{(\aa)}(t,x;s,y)\le \ff{C(t-s)}{\{\rr(x,y)\lor (t-s)^{\frac{1}{\aa}}\}^{d+\aa}},\\
& \big|\nn_x p_{b,c}^{(\aa)}(t,x; s,y)\big|\le \ff{C(t-s)^{\ff{\aa-1}\aa}}{\{\rr(x,y)\lor (t-s)^{\frac{1}{\aa}}\}^{d+\aa}},\ \ 0\le s<t,\ x,y\in M\end{split}\end{equation*}
 for some constant $C>1$, where $\rr$ is the Riemannian distance. The estimate of
 $\nabla_yp^{(\aa)}_{b,c}$ and the H\"older continuity of $\nn_x p_{b,c}^{(\aa)}$ are also considered.
The resulting estimates of the gradient and its H\"older continuity are new even in the standard case where $L=\DD$ on $\R^d$ and $b,c$ are time-independent. \end{abstract} \noindent

 \vskip 2cm

\section{Introduction}

In \cite{Zhang}, the two-sided Gaussian bounds were confirmed for the heat kernel of the time-dependent second order differential operator
${\rm div} (A\nn) + B\cdot \nn $ on $\R^d$, where $A: [0,\infty)\times\R^d\to\R^d\otimes\R^d$ is uniformly elliptic and uniformly H\"older continuous, and
$B: [0,\infty)\times\R^d\to\R^d$ is in a class of singular functions. In the same spirit, the sharp heat kernel bounds have been presented in \cite{BJ} for fractional Laplacian with perturbations. More precisely, let $\DD^{(\aa)}:=\DD^{\frac{\aa}{2}}$ be the fractional Laplacian on $\R^d$ for $\aa\in (1,2)$, and let $b: \R^d\to\R^d$ be in the Kato class $\K_d^{\alpha-1}$, i.e.
$$\lim_{\vv\downarrow 0} \sup_{x\in\R^d} \int_{|x-y|<\vv} \ff{|b(y)|}{|x-y|^{d+1-\aa}}\,\d y=0,$$
 or equivalently,
 $$\lim_{\vv\downarrow 0} \sup_{x\in\R^d} \int_{\R^d} \ff{|b(y)|(\vv\land |x-y|^\aa)}{|x-y|^{d+1}}\,\d y=0.$$
Then the heat kernel $p_b^{(\aa)} (t,x,y)$ of $\DD^{(\aa)} +\<b,\nn \cdot\>$ satisfies
\beq\label{1.1} \ff{t}{C(|x-y|\lor t^{\ff 1 \aa})^{d+\aa}}\le p_b^{(\aa)}(t,x,y)\le \ff{Ct}{(|x-y|\lor t^{\ff 1 \aa})^{d+\aa}}\end{equation}
for some constant $C>1.$
 Recently, this result was extended in \cite{CKS} to the Dirichlet heat kernel for  the fractional Laplacian with perturbations. The aim of this paper is to derive sharp heat kernel bounds for more general fractional diffusion operators with time-dependent perturbations, and to derive gradient estimates of the heat kernel which are new even in the framework of \cite{BJ}.

Let $M$ be a $d$-dimensional connected complete Riemannian manifold with Riemannian distance $\rr$. Let $L$ be an elliptic differential operator on $M$ generating a (sub-)Markov semigroup $P_t$. Then $P_t$ is a $C_0$-contraction semigroup on the Banach space $C_b(M)$ equipped with the uniform norm $\|\cdot\|_\infty,$ and $\D(L)\supset C_0^2(M)$, where $(L,\D(L))$ is the infinitesimal generator of $P_t$ on $C_b(M)$.

Throughout the paper, we assume that $P_t$ has a density $p(t,x,y)$ w.r.t. a reference measure $\mu$ on $M$ such that
\beq\label{1.2} \ff{\exp[-\ff{C\rr(x,y)^2}{t}]}{C(\mu(M)\land t)^{\ff d 2}}\le p(t,x,y)\le \ff{C\exp[-\ff{\rr(x,y)^2}{Ct}]}{(\mu(M)\land t)^{\ff d 2}},\ \ t>0, x,y\in M\end{equation}
and \beq\label{1.3} |\nn_x p(t,x,y)|\le \ff{C\exp[-\ff{\rr(x,y)^2}{Ct}]}{\ss t\,(\mu(M)\land t)^{\ff d 2} },\ \ t>0, x,y\in M \end{equation} hold for some constant  $C>1,$   where $\nn_x$ stands for the gradient operator w.r.t. variable $x$.  Consider the $\aa$-stable subordination of $P_t$:
$$
P_t^{(\aa)}:= \int_0^\infty P_s\,\mu_t^{(\aa)}(\d s),\ \ t\ge 0,
$$
where $\mu_t^{(\aa)}$ is a probability measure on $[0,\infty)$ with Laplace transform
$$\int_0^\infty \e^{-\ll s} \mu_t^{(\aa)}(\d s)=\e^{-t\ll^\aa},\ \ \ll\ge 0.$$ Then $P_t^{(\aa)}$ is  a $C_0$-contraction   semigroup on $C_b(M)$.  Let $L^{(\aa)}$ be  the infinitesimal generator of $P_t^{(\aa)}$. Then $\D(L^{(\aa)})\supset \D(L)\supset C_0^2(M)$, see e.g. \cite[Proposition 12.5]{SSV}.
By (\ref{1.2}), the density $p^{(\aa)}(t,x,y)$ of $P_t^{(\aa)}$ w.r.t. $\mu$ satisfies (see   Proposition \ref{PP0} below)
\beq\label{1.4} C^{-1}\xi^{(\aa)}(t,\rr(x,y)) \le p^{(\aa)}(t,x,y)\le C\xi^{(\aa)}(t,\rr(x,y)),\ \ t>0, x,y\in M\end{equation} for some constant $C>1$ and
\beq\label{1.4'} \xi^{(\aa)}(t,r):=\ff{t}{\mu(M)(r\lor t^{\ff 1\aa})^\aa}+ \ff{t}{(r\lor t^{\frac{1}{\aa}})^{d+\aa}},\ \ t>0, r\ge 0.\end{equation}

Now,   to make time-dependent first- and zero-order perturbations of $L^{(\aa)}$, let $b: [0,\infty)\times M\to TM$ and $c: [0,\infty)\times M\to \R$ be measurable. Consider
$$L_{b,c}^{(\aa)}(t,x) := L^{(\aa)}(x) +\< b(t,x), \nn_x \cdot\> + c(t,x),\ \ (t,x)\in [0,\infty)\times M.$$   To construct the heat kernel of this operator, we restrict   $|b|$ and $c$   in certain classes of functionals as in \cite{BJ,Zhang}.  To introduce these classes,   a function $f$ on $[0,\infty)\times M$ will be automatically extended to $\R\times M$ by letting $f(s,\cdot)=0$ for $s<0.$ For $\gg,\bb\ge 0$, define
$$K_{\aa,f}^{\gg,\bb}(\vv)= \sup_{(t,x)\in [0,\infty)\times M} \bigg\{\vv^{\ff \bb\aa}  \int^\vv_0\!\!\!\int_M  \ff{\xi^{(\aa)}(s,\rr(x,y))|f(t\pm s, y)|}{s^{\ff{\gg}\aa}(\vv-s)^{\ff\bb\aa}}\mu(\d y)\d s\bigg\},\ \ \vv>0.$$

\beg{defn} For $\gg,\bb\ge 0$, let
$$\KK_\aa^{\gg,\bb}=\KK_\aa^{\gg,\bb}(\mu)=\Big\{f\in\B(\R\times M):\ \lim_{\vv\downarrow 0} K_{\aa,f}^{\gg,\bb}(\vv)=0\Big\},$$where $\B(\R\times M)$ is the set of all measurable functions on $\R\times M.$ \end{defn}
It is easy to see that $\KK_\aa^{\gg,\bb}$ is decreasing in both
$\gg$ and $\bb$. According to Proposition \ref{PP0} below,
$\KK_\aa^{1,\bb}(\mu)\supset \K_d^{\alpha-1}(\mu)$ for any
$\bb\in [1,\aa)$, where $\K_d^{\alpha-1} (\mu)$ is the Kato
class on $M$ consisting of measurable functions $f$ with
\beq\label{KK} 1_{\{\mu(M)<\infty\}}\mu(|f|)<\infty,\ \
\lim_{\vv\downarrow 0} \sup_{x\in M} \int_{M} \ff{|f(y)|(\vv\land
\rr(x,y)^\aa)}{\rr(x,y)^{d+1}}\,\mu(\d y)=0.\end{equation}  When
$M=\R^d$ and $\mu(\d y)=\d y$, this class reduces back to the class
$\K_d^{\alpha-1}$ in \cite{BJ} as mentioned above. See also
Proposition \ref{PP2} for  explicit subclass of $\KK_\aa^{\gg,\bb}$
in the time-space functional space.

To introduce the heat kernel of $L_{b,c}^{(\aa)}$, let us look at the heat equation
\beq\label{Heat} \beg{cases} \pp_t u(t,s,\cdot)= L_{b,c}^{(\aa)} (t,\cdot) u(t,s,\cdot),\ \ t>s,\\
u(s,s,\cdot)=\varphi,\end{cases}\end{equation} where $s\ge 0$ and $\varphi\in C_b(M)$. Recall that $u$ is called a mild solution of this equation, if it satisfies
$$u(t,s,x)=P_{t-s}^{(\aa)} \varphi(x)+\int_s^t P_{t-r}^{(\aa)} \big\{\<b(r,\cdot), \nn u(r,s,\cdot)\> +c(r,\cdot)u(r,s,\cdot)\big\}(x)\d r,\ \ t\ge s.$$ Therefore, it is natural to construct the fundamental solution to the heat equation by solving the integral equation
\beq\label{1.6} \beg{split}  p_{b,c}^{(\aa)}(t,x;s,y)= &p^{(\aa)}(t-s, x,y)
 +\int_s^t \d r\int_M p^{(\aa)}(t-r, x,z)\\
 &\quad \cdot \big\{\<b(r,z),\nn_z p_{b,c}^{(\aa)}(r,z;s,y)\>
  + c(r,z) p_{b,c}^{(\aa)} (r,z; s,y)\big\}\mu(\d z)\end{split}\end{equation}for $t>s\ge 0, x,y\in M,$ so that the mild solution to (\ref{Heat}) can be formulated as
  $$u(t,s,\cdot)=P_{t,s}^{b,c}\varphi:= \int_M p_{b,c}^{(\aa)}(t,\cdot; s,y)\varphi(y) \mu(\d y).$$

We remark that following the argument of \cite{Zhang}, the heat kernel of  $\DD^{(\aa)}+\<b, \nn \cdot\>$ on $\R^d$ with  time-free $b$ was constructed in \cite{BJ} by solving the dual equation
\beq\label{1.7}p_b^{(\aa)}(t,x,y)= p^{(\aa)}(t,x,y)+\int_0^t\d r\int_{\R^d} p_b^{(\aa)}(t-r,x,z)\<b(z),\nn_z p^{(\aa)}(r,z,y)\>\d z,\end{equation}
where $p^{(\aa)}$ is the heat kernel for the $\aa$-stable operator $\DD^{(\aa)}.$ The advantage of (\ref{1.7}) is that it does not involve the  derivative of the unknown heat kernel, and hence easier to solve. On the other hand, the good point of   (\ref{1.6}) is that from which one can easily  derive the gradient estimate
and confirm the infinitesimal generator  of the solution.

The following three theorems are the main results of the paper.

\beg{thm}\label{T1.1} Assume $(\ref{1.2}), (\ref{1.3})$ and let $\aa\in (1,2).$
 If $|b|,c\in \KK_\aa^{1,1}$, then $(\ref{1.6})$ has a unique   solution $p_{b,c}^{(\aa)}(t,x;s,y)$ such that
 for all $t-s\in (0,1], x,y\in M$,
 \beq\label{A1}
 C^{-1}\xi^{(\aa)}(t-s,\rr(x,y))\le p_{b,c}^{(\aa)}(t,x;s,y)\le C \xi^{(\aa)}(t-s,\rr(x,y)), \end{equation} and
\beq\label{A2}|\nn_x p_{b,c}^{(\aa)}(t,x; s,y)|\le \ff{C\xi^{(\aa)}(t-s,\rr(x,y))} {(t-s)^{\ff{1}\aa}}\end{equation} hold for some constant $C>0$. Moreover, $p_{b,c}^{(\aa)}$ is continuous and satisfies the following two assertions: \beg{enumerate}
\item[$(1)$] For any $0\le s<r<t$ and $x,y\in M$,
$$p_{b,c}^{(\aa)}(t,x;s,y)= \int_M p_{b,c}^{(\aa)}(t,x;r,z)p_{b,c}^{(\aa)}(r,z;s,y)\mu(\d z);$$
\item[$(2)$] If  $\mu$ has a $C^1$-density w.r.t. the volume measure, $b\in C([0,\infty); L_{loc}^1(M\to TM;\mu)),c\in C([0,\infty); L_{loc}^1(M\to\R;\mu))$, then for any $\varphi,\psi\in C_0^2(M),$
$$\lim_{t\downarrow s} \ff 1 {t-s} \int_M \psi(P_{t,s}^{b,c}  \varphi -\varphi)\d\mu =\int_M\psi L_{b,c}^{(\aa)}(s,\cdot)\varphi\d\mu,\ \ s\ge 0.$$  \end{enumerate} \end{thm}
We remark that Theorem \ref{T1.1} not only generalizes the main result in \cite{BJ} for solution to (\ref{1.7}),   but also provide the new gradient estimate (\ref{A2}).
The next result says that under a Hessian upper bound condition of $p(t,x,y)$, we are able to further confirm the H\"older continuity estimate on $p_{b,c}^{(\aa)}$.
For $x\ne x'$, let   $\gg^{x,x'}: [0,1]\to M$ be the minimal geodesic from $x$ to $x'$, which might be non-unique if $x$ is in the cut-locus of $x'$. Define
\beq\label{ET} \eta(t;x,x';y)=p^{(\aa)}(t,x,y)+p^{(\aa)}(t,x',y)+\int^1_0p^{(\aa)}(t,\gamma^{x,x'}_\theta,y)\d\theta.\end{equation}

\beg{thm}\label{T1.2} Assume that $(\ref{1.2}), (\ref{1.3})$ and
\beq\label{HESS}\big|\nn^2_xp(t,x,y)\big|\le \ff{C\exp[-\ff{\rr(x,y)^2}{Ct}]}{t(\mu(M)\land t)^{\ff d 2}},\ \ t>0, x,y\in M\end{equation}hold for some constant $C>1$.
Let $\aa\in (1,2).$ If $|b|,c\in \KK_\aa^{\bb,\bb}$ for some $\bb\in (1,\aa)$, then there exists a constant $C'>0$ such that
$$\Big|\nn_x p_{b,c}^{(\aa)}(t,x; s,y)- \parallel_{x'\to x} \nn_{x'} p_{b,c}^{(\aa)}(t,x'; s,y)\Big|\le \ff{C'\rr(x,x')^{\bb-1}\eta(t-s;x,x';y)}
{(1\land (t-s))^{\ff\bb\aa}}$$ holds for all $0\le s<t$ and $x,x',y\in M$, where $\parallel_{x'\to x}$ denotes
the parallel transport along the geodesic $\gamma^{x,x'}$.   \end{thm}

Finally, we consider   the derivative estimate of $p_{b,c}^{(\aa)}$ w.r.t. the variable ``$y$''.

\beg{thm}\label{T1.3}
In addition to the assumptions of Theorem $\ref{T1.1}$, we also assume that
$p^{(\alpha)}(t,x,y)=p^{(\alpha)}(t,y,x)$ and $\mathrm{div}_\mu b\in\KK^{1,1}_\alpha$ exists, where $\div_\mu b(s,\cdot)$ is the unique $($if exists$)$ element in $L^1_{loc}(M\to\R;\mu)$ such that
$$\int_M (\div_\mu b(s,\cdot)) f\d\mu= \int_M \<b(s,\cdot),\nn f\> \d\mu,\ \ f\in C_0^1(M).$$  Then
\beq\label{A22}
|\nn_y p_{b,c}^{(\aa)}(t,x; s,y)|\le \ff{C\xi^{(\aa)}(t-s,\rr(x,y))} {(t-s)^{\ff{1}\aa}}
\end{equation}
hold for some constant $C>0$.
\end{thm}
The remainder of the paper is organized as follows. We present in Section 2 some estimates on $p^{(\aa)}(t,x,y)$
and characterization of  the class $\KK_\aa^{\gg,\bb}$, then prove the above theorems in Section 3.
Finally, some examples are presented in Section 4 to illustrate the above three theorems.

\section{Some preliminaries}

In this section we aim to characterize the class $\KK_\aa^{\gg,\bb}$ and to present some estimates on $p^{(\aa)}$ which will be used in the proofs of Theorems \ref{T1.1} and \ref{T1.2}.

\beg{prp}\label{PP0} $\KK_{\aa}^{1,\bb}(\mu)\supset
\K_d^{\alpha-1}(\mu)$ holds for $\bb\in [1,\aa)$, where
$\K_d^{\alpha-1}(\mu)$ is fixed by $(\ref{KK})$.
\end{prp}
\beg{proof} Since
$$\int_0^\vv\Big\{ \big(s^{1-\ff 1\aa}r^{-(d+\aa)}\big)\land s^{-\ff{d+1}\aa}\Big\}\d s\le C \big\{r^{\aa-d-1}\land (\vv r^{-(d-1)})\big\},\ \ \vv,r>0$$ holds for some constant $C>0$, it is easy to see that
\beg{equation*}\beg{split} &\vv^{\ff\bb \aa} \int_0^{\ff\vv 2} \int_M |f(y)| \xi^{(\aa)}(s,\rr(x,y))s^{-\ff 1 \aa}(\vv-s)^{-\ff\bb\aa}\mu(\d y)\d s\\
&\le C_1 \vv^{\ff{\aa-1}\aa}\mu(|f|)1_{\{\mu(M)<\infty\}} +C_1 \int_M |f(y)| \mu(\d y) \int_0^{\ff\vv 2}
 \Big\{ \big(s^{1-\ff 1\aa}r^{-(d+\aa)}\big)\land s^{-\ff{d+1}\aa}\Big\} \d s\\
&\le C_1 \vv^{\ff{\aa-1}\aa}\mu(|f|)1_{\{\mu(M)<\infty\}} +C_2 \int_{M} \ff{|f(y)|(\vv\land \rr(x,y)^\aa)}{\rr(x,y)^{d+1}}\,\mu(\d y)\end{split}\end{equation*} holds for some constants $C_1,C_2>0.$ Similarly, the same estimate holds for $$\vv^{\ff\bb \aa} \int_{\ff \vv 2}^{\vv} \int_M |f(y)| \xi^{(\aa)}(s,\rr(x,y))s^{-\ff 1 \aa}(\vv-s)^{-\ff\bb\aa}\mu(\d y)\d s.$$ Therefore, the proof is finished. \end{proof}

In the next result, we present a lower bound of $\KK_{\aa}^{\gg,\bb}(\mu)$ in the class of time-space functions.

\beg{prp}\label{PP2} Assume that
 \beq\label{EI3}\mu(B(x,s))\le C s^{d},\ \ s\ge 0, x\in M\end{equation} holds for some constant $C>0$.
Let $\alpha\in(0,2)$, $\gamma,\beta\in[0,\alpha)$ and $p,q\in[1,\infty]$. If
\beq\label{EI4}
\frac{d}{p}+\frac{\alpha}{q}<\alpha-\gamma,\ \ q>\frac{\alpha}{\alpha-\beta},\end{equation}
then
$$
L^q(\R;L^p(M, \mu))\subset \KK_\aa^{\gamma,\beta}(\mu).
$$
\end{prp}
\begin{proof}
Using H\"older's inequality, it is enough to prove
\beq\label{EW} \lim_{t\downarrow 0} \sup_{x\in M} I(x,t)=0,\end{equation} where
$$
I(x,t):=t^{\frac{\beta q^*}{\alpha}}\int^t_0\left(\int_M\xi^{(\alpha)}(s,\rr(x,y))^{p^*}\mu(\d  y)\right)^{\frac{q^*}{p^*}}
s^{-\frac{\gamma q^*}{\alpha}}(t-s)^{-\frac{\beta q^*}{\alpha}}\d  s
$$
for $q^*=\frac{q}{q-1}$ and $p^*=\frac{p}{p-1}$. By the definition of $\xi^{(\alpha)}$, there exists a constant $C_1>0$ such that
\begin{align*}
&\int_M\xi^{(\alpha)}(s,\rr(x,y))^{p^*}\mu(\d  y)\\&\le C_1\bigg\{1+ \int_{B(x,s^{\frac{1}{\alpha}})}s^{-\frac{dp^*}{\alpha}}\mu(\d  y)
+s^{p^*}\int_{B(x,s^{\frac{1}{\alpha}})^c}\rr(x,y)^{-(d+\alpha)p^*}\mu(\d  y)\bigg\}\\
&=:C_1\big\{1+J_1(x,s)+J_2(x,s)\big\},\ \ s\in (0,1].
\end{align*}
By (\ref{EI3}), there exists a constant $C>0$ such that
$
1+J_1(x,s)\leq   Cs^{\frac{d-dp^*}{\alpha}}
$
and
\begin{align*}
J_2(x,s)&= s^{p^*}\sum_{n=0}^\infty\int_{B(x,2^{n+1}s^{\frac{1}{\alpha}})-B(x,2^ns^{\frac{1}{\alpha}})}\rr(x,y)^{-(d+\alpha)p^*}\mu(\d  y)\\
&\leq s^{p^*-\frac{(d+\alpha)p^*}{\alpha}}\sum_{n=0}^\infty2^{-n(d+\alpha)p^*}\mu(B(x,2^{n+1}s^{\frac{1}{\alpha}}))\\
&\leq C s^{ -\frac{dp^*}{\alpha}}\sum_{n=0}^\infty2^{-n(d+\alpha)p^*}2^{(n+1)d}s^{\ff d\aa} =  Cs^{\ff{d-dp^*}\aa}
\end{align*} hold for $s\in (0,1].$
Thus,
$$
I(x,t)\leq C_2 t^{\frac{\beta q^*}{\alpha}}\int^t_0s^{\frac{dq^*}{\alpha p^*}-\frac{dq^*}{\alpha}
-\frac{\gamma q^*}{\alpha}}(t-s)^{-\frac{\beta q^*}{\alpha}}\d  s
=C_2 t^{1+\theta}\int^1_0 s^\theta (1-s)^{-\frac{\beta q^*}{\alpha}}\d  s\leq C_3t^{1+\theta},
$$ holds for some constants $C_2,C_3>0,$ and all $t\in (0,1]$,
where $\theta=\frac{dq^*}{\alpha p^*}-\frac{dq^*}{\alpha}-\frac{\gamma q^*}{\alpha}>-1$
and $\frac{\beta q^*}{\alpha}<1$  by (\ref{EI4}). Then (\ref{EW}) holds.
\end{proof}

Next, we consider estimates on $p^{(\aa)}(t,x,y)$.

\beg{prp}\label{PP3} Assume $(\ref{1.2})$.  \beg{enumerate} \item[$(1)$]     $(\ref{1.4})$ holds for some   constant $C>1.$
\item[$(2)$] If there exist constants   $C_1,C_2>0$ and a natural number $k\ge 1$ such that
\beq\label{DD} |\nn^i_x p(t,x,y)|\le \ff{C_1\exp[-\ff{C_2\rr(x,y)^2}t]}{t^{\ff i 2}(\mu(M)\land t)^{\ff d 2}},\ \ t>0, x,y\in M, 0\le i\le k,\end{equation}
then \beq\label{1.2'} |\nn^k_x p^{(\aa)}(t,x,y)|\le \ff{C\xi^{(\aa)}(t,\rr(x,y))}{t^{\ff k \aa}},\ \ t>0, x,y\in M\end{equation} holds for some constant $C>0.$
\item[$(3)$] If $(\ref{DD})$ holds for $k=1,2$, then for any $\bb\in (1,\aa)$ there exists a constant $C>0$ such that
$$\Big| \nn_x p^{(\aa)}(t,x,y)- \parallel_{x'\to x}\nn_{x'} p^{(\aa)}(t,x',y)\Big|\le
C t^{-\frac{\beta}{\alpha}}\rr(x,x')^{\beta-1}\eta(t;x,x';y),$$ where $\eta$ is in $(\ref{ET})$. \end{enumerate}
 \end{prp}

\beg{proof} (1) According to the proof of \cite[Theorem 3.1]{BSS}, for any $\ll>0$ and $m\ge 0$ there exists a constant $C_1>1$ such that
\beq\label{BJ} \ff{t}{C_1 (r\lor t^{\ff 1 \aa})^{m+\aa}}\le \int_0^\infty \ff{\exp[-\ff{\ll r^2}s] }{s^{\ff m 2}}\mu_t^{(\aa)}(\d s) \le
\ff{C_1t}{ (r\lor t^{\ff 1 \aa})^{m+\aa}}\end{equation} holds for any $r,t>0.$ Combining this with the second inequality in (\ref{1.2}) we obtain
\beg{equation*}\beg{split} p^{(\aa)}(t,x,y)&\le   C \int_0^\infty \ff{\exp[-\ff{  \rr(x,y)^2}{Cs}] }{s^{\ff d 2}}\mu_t^{(\aa)}(\d s)
+ \ff{C}{\mu(M)}1_{\{\mu(M)<\infty\}}\int_0^\infty\e^{-\ff{\rr(x,y)^2}{Cs}}\mu_t^{(\aa)}(\d s)\\
&\le     \ff {C't}{(\rr(x,y)\lor t^{\ff 1 \aa})^{(d+\aa)}} + \ff {C't}{(\rr(x,y)\lor t^{\ff 1 \aa})^{\aa}} = C'\xi^{(\aa)}(t,\rr(x,y))\end{split}\end{equation*} for some constant $C'>1.$  On the other hand, noting that
$$\ff 1{(\mu(M)\land t)^{\ff d 2}}\ge \ff 1 2 \Big(\ff 1 {t^{\ff d 2}}+ \ff{1_{\{\mu(M)<\infty\}}} {\mu(M)^{\ff d 2}}\Big),$$
we obtain the desired lower bound estimate by using the first inequality  in (\ref{1.2}).

(2) It is well known that (cf. (14) in \cite{BSS})
\begin{align}
\mu_t^{(\aa)}(\d s)\le C_0 t s^{-\ff{2+\aa}2} \exp[-ts^{-\ff \aa 2}]\d s\label{QQ}
\end{align}
holds for some constant $C_0>0$. Then (\ref{DD}) yields that
$$\sup_{x,y}\sup_{0\le i\le k}|\nn^i_x p(\cdot,x,y)|\in L^1(\mu_t^{(\aa)}),$$ so that by the dominated convergence theorem we obtain   \beg{equation*}\beg{split}
&|\nn^k_x p^{(\aa)}(t,x,y)| \\
&\le C_0C_1 t\int_0^\infty \big\{s^{-\ff{d+\aa+k+2}2}
+1_{\{\mu(M)<\infty\}} s^{-\ff{\aa+k+2}2}\big\}\exp\Big[-\ff{t}{s^{\ff\aa 2}}-\ff{C_2\rr(x,y)^2} s\Big]\d s\\
& \le C_0C_1(I_1\land I_2),\end{split}\end{equation*}
where
\beg{equation*}\beg{split} I_1&:=   t\int_0^\infty  \big\{s^{-\ff{d+\aa+k+2}2}
+1_{\{\mu(M)<\infty\}} s^{-\ff{\aa+k+2}2}\big\}\exp\Big[-\ff{C_2\rr(x,y)^2} s\Big]\d s \\
&\le C_3 t\bigg\{\rr(x,y)^{-(d+\aa+k)}\int_0^\infty r^{\ff{d+\aa+k-2}2}\e^{-r}\d r +1_{\{\mu(M)<\infty\}}\rr(x,y)^{-(\aa+k)}\int_0^\infty r^{\ff{\aa+k-2}2}\e^{-r}\d r\bigg\}\\
&\le C_4 t\big\{\rr(x,y)^{-(d+\aa+k)}+1_{\{\mu(M)<\infty\}} \rr(x,y)^{-(\aa+k)}\big\},\end{split}\end{equation*} and
\beg{equation*}\beg{split} I_2&:=   t\int_0^\infty  \big\{s^{-\ff{d+\aa+k+2}2}
+1_{\{\mu(M)<\infty\}} s^{-\ff{\aa+k+2}2}\big\}\exp\Big[-\ff{t}{s^{\ff\aa 2}}\Big]\d s \\
&\le C_3 \bigg\{t^{-\ff{d+k}\aa}  \int_0^\infty r^{\ff{d+k}\aa}\e^{-r}\d r +1_{\{\mu(M)<\infty\}}t^{-\ff k\aa}\int_0^\infty r^{\ff k \aa}\e^{-r}\d r\bigg\}\\
&\le C_4 \big\{t^{-\ff{d+k}\aa}+1_{\{\mu(M)<\infty\}} t^{-\ff k\aa}\big\}\end{split}\end{equation*}
for some constants $C_3,C_4>0.$ Therefore,
\beq\label{DD1}\beg{split}&|\nn^k_x p^{(\aa)}(t,x,y)|\\
&\le C_4\min\Big\{t\rr(x,y)^{-(d+\aa+k)}+t1_{\{\mu(M)<\infty\}} \rr(x,y)^{-(\aa+k)},t^{-\ff{d+k}\aa}+1_{\{\mu(M)<\infty\}} t^{-\ff k\aa}\Big\}.\end{split}\end{equation}
From this we complete the proof by considering the following two cases respectively.

(i) If $\rr(x,y)\le t^{\ff 1 \aa}$ then (\ref{DD1}) implies
$$|\nn^k_x p^{(\aa)}(t,x,y)|
\le C_4\big\{t^{-\ff{d+k}\aa}+1_{\{\mu(M)<\infty\}} t^{-\ff k\aa}\big\}.$$
Moreover, in this case (\ref{1.4'}) implies
$$\xi^{(\aa)}(t,\rr(x,y))\ge \ff 1 {\mu(M)} + t^{-\ff d \aa}.$$ Therefore, (\ref{1.2'}) holds for some constant $C.$

(ii) If $\rr(x,y)\ge t^{\ff 1\aa}$, then from (\ref{DD1}) and (\ref{1.4'}) we obtain
\beg{equation*}\beg{split} |\nn^k_x p^{(\aa)}(t,x,y)|
&\le C_4\big\{t\rr(x,y)^{-(d+\aa+k)}+t1_{\{\mu(M)<\infty\}} \rr(x,y)^{-(\aa+k)}\big\}\\
&\le \ff{C}{t^{\ff k\aa}} \xi^{(\aa)}(t,\rr(x,y))\end{split}\end{equation*}   for some constant $C>0.$

(3) Since (\ref{DD}) implies (\ref{1.2'}) for $k=1,2$, we have
\begin{align*}
\Big| \nn_x p^{(\aa)}(t,x,y)- \parallel_{x'\to x}\nn_{x'} p^{(\aa)}(t,x',y)\Big|&\le
\rr(x,x')\int^1_0|\nabla^2_xp^{(\alpha)}(t,\gamma^{x,x'}_\theta,y)|\d\theta\\
&\leq Ct^{-\frac{2}{\alpha}}\rr(x,x')\int^1_0|p^{(\alpha)}(t,\gamma^{x,x'}_\theta,y)|\d\theta.
\end{align*}
Hence, by (\ref{1.2'}) for $k=1$ and Young's inequality, we obtain
\begin{align*}
&\Big| \nn_x p^{(\aa)}(t,x,y)- \parallel_{x'\to x}\nn_{x'} p^{(\aa)}(t,x',y)\Big|\\
&\le\Big(|\nabla_x p^{(\alpha)}(t,x,y)|^{2-\beta}+|\nabla_{x'} p^{(\alpha)}(t,x',y)|^{2-\beta}\Big)\\
&\qquad\times |\nabla_x  p^{(\alpha)}(t,x,y)-\parallel_{x'\to x}\nabla_{x'} p^{(\alpha)}(t,x',y)|^{\beta-1}\\
&\le C_3 t^{-\ff\bb\aa}\rr(x,x')^{\bb-1} \big\{p^{(\aa)}(t,x,y)+p^{(\aa)}(t,x',y)\big\}^{2-\bb} \\
&\qquad\times\bigg(\int^1_0|p^{(\alpha)}(t,\gamma^{x,x'}_\theta,y)|\d\theta\bigg)^{\bb-1}\\
&\leq C_4 t^{-\frac{\beta}{\alpha}}\rr(x,x')^{\beta-1}\eta(t;x,x';y)
\end{align*}for some constants $C_3,C_4>0.$  Then the proof is finished.
\end{proof}

Finally, we present below a (3P)-inequality as in \cite[Theorem 4]{BJ}.

\beg{prp}\label{P3P} There exists a constant $C>0$ such that
\beq\label{3P0} \xi^{(\aa)}(t,r)\land \xi^{(\aa)}(s,u)\le C \xi^{(\aa)}(t+s, r+u),\ \ s,t,r,u>0.\end{equation} Consequently, there exists a constant $C>0$ such that  for $s,t>0, x,y,z\in M$,
\beq\label{3P} p^{(\aa)}(t,x,z)p^{(\aa)}(s,z,y)\le Cp^{(\aa)}(s+t,x,y)(p^{(\aa)}(t,x,z)+p^{(\aa)}(s,z,y)).\end{equation} \end{prp}

\beg{proof} (1) According to the proof of \cite[Theorem 4]{BJ}, for any $m\ge 0$ the function $\xi_m(t,r):= t^{-\ff m \aa}\land (tr^{-(m+\aa)})$ satisfies
\beq\label{3P'} \xi_m(t,r)\land \xi_m(s,u)\le 2^{\ff {6m}\aa} \xi_m(t+s, r+u),\ \ t,r,s,u>0.\end{equation} So, it suffices to prove (\ref{3P0}) for $\mu(M)<\infty$. In this case the proof of (\ref{3P0}) can be finished by considering the following two situations.

(i) If either $r\lor t, s\lor u\ge 1$ or $r\lor t,s\lor u<1,$ then for   $m=d$ or $m=0$ respectively one derives from (\ref{3P'})  that
\beg{equation*}\beg{split}& \xi^{(\aa)}(t,r)\land \xi^{(\aa)}(s,u) \le \Big(1+\ff 1 {\mu(M)}\Big) \big\{\xi_m(t,r)\land \xi_m(s,u)\big\}\\
&\le \ff{2^{6m}(1+\mu(M))}{\mu(M)}  \xi_m(t+s,r+u)\le \ff{2^{6m}(1+\mu(M))^2}{\mu(M)} \xi^{(\aa)}(t+s,r+u).\end{split}\end{equation*}

(ii) If e.g. $r\lor t\le 1$ but $s\lor u>1$, then
\beg{equation*}\beg{split} &\xi^{(\aa)}(t,r)\land \xi^{(\aa)}(s,u)\le \xi^{(\aa)}(s,u) \le \ff{s(1+\mu(M))}{\mu(M) (u\lor s^{\ff 1\aa})^{\aa}}\\
&\le  \ff{(1+\mu(M))(t+s)}{\mu(M) (u^\aa\lor s) }\le \ff{2^\aa(1+\mu(M))(t+s)}{\mu(M) \{(r+u)^\aa\lor (t+s)\} }\\
&\le 2^\aa(1+\mu(M))\xi^{(\aa)}(t+s,r+u).\end{split}\end{equation*}

(2) Combining (\ref{3P0}) with the inequality $ab\le (a+b)(a\land b), a,b>0,$ we obtain
$$\xi^{(\aa)}(t,r)\xi^{(\aa)}(s,u)\le C  \xi^{(\aa)}(t+s, r+u)(\xi^{(\aa)}(t,r)+\xi^{(\aa)}(s,u)).$$ Since $\xi^{(\aa)}(t,r)$ is decreasing in $r$, combining this with Proposition \ref{PP3} (1) and noting that $\rr(x,z)+\rr(z,y)\ge\rr(x,y)$, we prove (\ref{3P}) for some (different) constant $C>0.$
\end{proof}

\section{Proofs of Theorems \ref{T1.1}, \ref{T1.2} and \ref{T1.3}}
To construct the solution of (\ref{1.6}), we make use of the argument of Picard iteration as in \cite{Zhang, BJ}. For $t>s\ge 0$ and $x,y\in M,$ let $p_0(t,x;s,y)=p^{(\aa)}(t-s,x,y)$ and
\beg{equation}\beg{split}  \label{ET1}
 p_n(t,x;s,y)&=p^{(\aa)}(t-s,x,y)  +\int^t_s\!\!\!\int_Mp^{(\aa)}(t-r,x,z)\\
& \quad \cdot \Big\{\big\<b(r,z), \nabla_z  p_{n-1}(r,z;s,y)\>
 +  c(r,z)p_{n-1}(r,z;s,y)\Big\}\mu(\d z) \d r
\end{split}\end{equation} for $n\ge 1$.
Moreover, let
$\Theta_0(t,x;s,y):=p^{(\aa)}(t-s,x,y)$ and
$$
\Theta_n(t,x;s,y):=p_n(t,x;s,y)-p_{n-1}(t,x;s,y),\ \ n\ge 1.
$$
 It is clear that
\beq\label{2.3}\beg{split}
\Theta_n(t,x;s,y)&=\int^t_s\!\!\!\int_Mp^{(\aa)}(t-r,x,z)\<b(r,z), \nabla_z  \Theta_{n-1}(r,z;s,y)\>\mu(\d z) \d r \\
&\quad+\int^t_s\!\!\!\int_Mp^{(\aa)}(t-r,x,z) c(r,z)\Theta_{n-1}(r,z;s,y)\mu(\d z) \d r.\end{split}\end{equation}

\beg{lem}\label{L3.1} Assume $(\ref{1.2}), (\ref{1.3})$ and let $|b|,c\in \mathbb K_\aa^{1,1}.$ Let
$$\ell(r)=\sup_{\vv\in (0, r]} \big\{K_{\aa,|b|}^{1,1}(\vv)+K_{\aa,c}^{1,1}(\vv)\big\},\ \ r>0.$$ Then there exists a constant $c_0>0$ such that for any $n\ge 0,$ $p_n$ (hence $\Theta_n)$ is well defined  and
\beg{equation}\beg{split}\label{ES} &|\nabla_x \Theta_n(t,x;s,y)| \leq \{c_0\ell(t-s)\}^n(t-s)^{-\frac{1}{\alpha}}p^{(\aa)}(t-s,x,y),\\
&|\Theta_n(t,x;s,y)| \leq \{c_0\ell(t-s)\}^np^{(\aa)}(t-s,x,y), \ \ t>s\ge 0, x,y\in M.\end{split}\end{equation}\end{lem}

\begin{proof} According to Propositions \ref{PP3} and \ref{P3P}, we may take a constant   $C\ge 1$ such that (\ref{1.4}), (\ref{1.2'}) for $k=1,$ and (\ref{3P}) hold. Take $c_0=4C.$ Then the assertion holds for $n=0$. Assume it holds for $n\le m$ for some $m\ge 0$, then it is easy to see from (\ref{1.4}) and $|b|,c\in \KK_\aa^{1,1}$ that $p_{m+1}$ is well defined. It remains to prove (\ref{ES}) for $n=m+1$.
For any unit vector $U\in T_xM,$   let $x_\vv=\exp[\vv U], \vv\ge 0.$   According to the assertion for  $n=m$, it follows from (\ref{3P}) that
\beg{equation*}\beg{split} &\ff 1 \vv |\Theta_{m+1}(t, x_\vv;s,y)-\Theta_{m+1} (t,x;s,y)|\\
&\le  \ff 1 \vv \int_0^\vv\d\theta \int^t_s\!\!\!\int_M|\nn_x p^{(\aa)}(t-r,x_\theta, z)|\\
&\qquad\qquad\cdot \Big\{|b(r,z)|\cdot|\nn_z \Theta_{m}(r,z;s,y)|
 + |c(r,z)| \Theta_m(r, z; s,y)\Big\}\mu(\d z)\d r\\
&\le \ff {(c_0\ell(t-s))^m}\vv \int_0^\vv\d\theta \int^t_s\!\!\!\int_M (t-r)^{-\ff 1\aa}p^{(\aa)}(t-r,x_\theta,z)p^{(\aa)}(r-s,z,y)\\
&\qquad\qquad\cdot \Big\{|b(r,z)|(r-s)^{-\ff 1\aa}+|c(r,z)|\Big\}\mu(\d z)\d r\\
&\le \ff{C(c_0\ell(t-s))^m}{\vv }
   \int_0^\vv p^{(\aa)}(t-s, x_\theta,y)\d\theta\\
  &\qquad  \int^t_s\!\!\!\int_M\Big\{|b(r,z)|\big(p^{(\aa)}(t-r, x_\theta, z)+ p^{(\aa)}(r-s,z,y)\big) (t-r)^{-\ff 1 \aa} (r-s)^{-\ff 1\aa}\\
&\qquad\qquad  + |c(r,z)|\big(p^{(\aa)}(t-r,x_\theta,z)+p^{(\aa)}(r-s,z,y)\big) (t-r)^{-\ff 1\aa}\Big\}\mu(\d z)\d r\\
&\le C (c_0\ell(t-s))^m \big(2K_{\aa,|b|}^{1,1}+K_{\aa,c}^{1,0}+K_{\aa,c}^{0,1}\big)(t-s)\bigg(\ff 1\vv  \int_0^\vv p^{(\aa)}(t-s, x_\theta,y)\d\theta\bigg)\\
&\le (c_0\ell(t-s))^{m+1}\bigg(\ff 1\vv  \int_0^\vv p^{(\aa)}(t-s, x_\theta,y)\d\theta\bigg)\to (c_0\ell(t-s))^{m+1}p^{(\aa)}(t-s,x,y) \end{split}\end{equation*} as $\vv\to 0.$
Then the first inequality in (\ref{ES}) holds for $n=m+1$. Similarly, and even simpler, we have
$$|\Theta_{m+1}(t,x;s,y)| \leq \{c_0\ell(t-s)\}^{m+1}p^{(\aa)}(t-s,x,y).$$ Therefore, the proof is finished. \end{proof}

\beg{proof}[Proof of Theorem \ref{T1.1}] It is standard that we   need  only  to find $t_0>0$ such that  (\ref{1.6}) has a unique solution satisfying (\ref{A1}) and (\ref{A2}), and to verify assertion (1) with $t-s\le t_0$   and assertion  (2). Let $t_0>0$ be such that $c_0\ell(t_0)\le \ff 1 3$,
where $\ell$ is defined in Lemma \ref{L3.1}.

(a) Construction of the solution.  Define
$$p_{b,c}^{(\aa)}(t,x;s,y) =\sum_{n=0}^\infty \Theta_n(t,x;s,y),\ \ t-s\in (0,t_0], x,y\in M.$$ By Lemma \ref{L3.1}, this series, as well as $ \sum_{n=0}^\infty \nn_x  \Theta_n(t,x;s,y)$,
converge  uniformly on $\{(t,x;s,y):\ t-s\in (0,t_0], x,y\in M\}.$ Then (\ref{A2}) holds. By letting $n\to\infty$ in (\ref{ET1}), we see that  (\ref{1.6}) holds. Moreover, since all $\Theta_n$ are jointly continuous, so is $p_{b,c}^{(\aa)}$.  Finally, by Lemma \ref{L3.1}, we have
\beg{equation*}\beg{split} &|p_{b,c}^{(\aa)}(t,x;s,y)-p^{(\aa)}(t-s,x,y)|\le \sum_{n=1}^\infty |\Theta_n(t,x;s,y)|\\
&\le \ff{c_0\ell(t-s)}{1-c_0\ell(t-s)}p^{(\aa)}(t-s,x,y)\le \ff 1 2 p^{(\aa)}(t-s,x,y).\end{split}\end{equation*} Then (\ref{A1}) follows from (\ref{1.4})   ensured   by Proposition \ref{PP3}(1).

(b) Uniqueness. Let $\tt p^{(\aa)}_{b,c}$ be another solution to (\ref{1.6}) satisfying (\ref{A2}). Then the induction argument in the proof of Lemma \ref{L3.1} implies that
$\Theta:=p_{b,c}^{(\aa)}- \tt p_{b,c}^{(\aa)}$ satisfies $|\Theta(t,x;s,y)|\le (c_0\ell(t-s))^n p^{(\aa)}(t-s,x,y)$ for all $n\ge 0$, so that letting $n\to\infty$ we derive $\Theta(t,x;s,y)=0$ for $t-s\le t_0.$  Thus, the solution is unique.

(c) For (1) it suffices to prove that for any $\varphi\in C^\infty_0( M)$,
\begin{align}
P^{b,c}_{t,s}\varphi(x)=P^{b,c}_{t,r}P^{b,c}_{r,s}\varphi(x),\ \ s\leq r\leq t,\label{EU2}
\end{align}
where
$$
P^{b,c}_{t,s}\varphi(x):=\int_Mp^{(\aa)}_{b,c}(t,x;s,y)\varphi(y)\mu(\d y).
$$
 Set
$$
g_{r,s}(x)=\big\<b(r,x),\nabla P^{b,c}_{r,s} \varphi(x)\big\>+c(r,x)P^{b,c}_{r,s}\varphi(x),\ \ r>s.
$$
By (\ref{1.6}) we have
\beg{equation*}\beg{split}
P^{b,c}_{t,s}\varphi(x)&=P_{t-s}^{(\aa)}\varphi(x)+\int^t_sP_{t-r'}^{(\aa)}g_{r',s}(x)\d r'  \\
&=P_{t-r}^{(\aa)}P_{r-s}^{(\aa)}\varphi(x)+\int^r_sP_{t-r}^{(\aa)}P_{r-r'}^{(\aa)}g_{r',s}(x)\d r'
+\int^{t}_rP_{t-r'}^{(\aa)}g_{r',s}(x)\d r \\
&=P_{t-r}^{(\aa)}P^{b,c}_{r,s}\varphi(x)+\int^t_rP_{t-r'}^{(\aa)}\Big(\<b(r',\cdot),\nabla P^{b,c}_{r',s} \varphi\>\Big)(x)\d r'\\
&\qquad+\int^t_rP_{t-r'}^{(\aa)}\Big(c(r',\cdot)P^{b,c}_{r',s}\varphi\Big)(x)\d r'. \end{split}\end{equation*}
On the other hand,
\begin{align*}
P^{b,c}_{t,r}P^{b,c}_{r,s}\varphi(x)&=P^{(\aa)}_{t-r}P^{b,c}_{r,s}\varphi(x)
+\int^t_rP^{(\aa)}_{t-r'}\Big(\<b(r',\cdot),\nabla (P^{b,c}_{r',r}P^{b,c}_{r,s} \varphi)\>\Big)(x)\d r'\\
&\qquad+\int^t_rP^{(\aa)}_{t-r'}\Big(c(r',\cdot)P^{b,c}_{r',r}P^{b,c}_{r,s} \varphi\Big)(x)\d r'.
\end{align*}
By the uniqueness as observed in (b), we obtain (\ref{EU2}).

(d)   Finally, we prove (2). Let $\varphi,\psi\in C_0^2(M)$. By (\ref{1.6}), we have
\beq\label{WE} \beg{split} &\ff{P_{t,s}^{b,c}\varphi-\varphi}{t-s} - L_{b,c}^{(\aa)}(s)\varphi\\
&= \Big(\ff{P_{t-s}^{(\aa)}\varphi -\varphi}{t-s} -L^{(\aa)}\varphi\Big) + \ff 1 {t-s} \int_s^t\<b(r,\cdot)-b(s,\cdot), \nn P_{r,s}^{b,c}\varphi\>\d r\\
&\quad+\ff 1{t-s} \int_s^t \Big\{(c(r,\cdot)-c(s,\cdot))P_{r,s}^{b,c}\varphi +c(s,\cdot)(P_{r,s}^{b,c}\varphi -\varphi)\Big\}\d r\\
 &\quad+\ff 1 {t-s} \int_s^t \<b(s,\cdot), \nn P^{b,c}_{r,s}\varphi -\nn\varphi\>\d r\\
&=: I_0(t,s) +I_1(t,s)+ I_2(T,s)+I_3(t,s).\end{split}\end{equation}
Since $\varphi\in C_0^2(M)\subset \D(L^{(\aa)}),$ we have
\beq\label{WE2} \lim_{t\downarrow s} \|I_0(t,s)\|_\infty =0,\ \ s\ge 0.\end{equation}
Fix $s\geq 0$ and set
$$
u(t,x):=P^{b,c}_{t,s}\varphi(x),\ \ t\ge s.
$$ First of all, by (\ref{1.4}), (\ref{A1}) and the contraction of $P_t^{(\aa)}$ we have
\beq\label{FF0} \sup_{t\in [s, 1+s]} \|u(t,\cdot)\|_\infty  <\infty.\end{equation}
Next, by (\ref{1.3}) and Proposition \ref{PP3}(2),   (\ref{1.2'}) holds for $k=1$. Combining this with  (\ref{1.4}), we obtain
\beq\label{FF1}\beg{split}
\|\nabla P_t^{(\aa)}\varphi-\nabla\varphi\|_\infty&\leq\int^t_0\|\nabla P^{(\alpha)}_s L^{(\alpha)}\varphi\|_\infty\d s\\
&\leq C_1\|L^{(\alpha)}\varphi\|_\infty\int^t_0s^{-\frac{1}{\alpha}}\d s\leq C,\ \ t\in [s,1+s]
\end{split}\end{equation} for some constants $C_1,C>0.$
Next, let
$$\Theta_{t,s}^{(n)}\varphi  = \int_M \Theta_n(t,\cdot;s,y)\varphi(y)\mu(\d y),\ \ t\ge s.$$
By (\ref{ES}), (\ref{1.4}) and noting that $\|P^{(\aa)}_r \varphi\|_\infty\le \|\varphi\|_\infty <\infty, r\ge 0,$ we obtain
\beg{equation*}\beg{split} &\|\nn \Theta_{t,s}^{(n)} \varphi\|_\infty \\
&\le C\|\varphi\|_\infty (c_0\ell(t-s))^{n-1}\bigg\|\int^t_s\!\!\!\int_M \xi^{(\aa)}(t-r,\rr(\cdot,z))\Big(\ff{|b(r,z)|}{(t-r)^{\ff 1 \aa}}+|c(r,z)|\Big)\mu(\d z)\d r \bigg\|_\infty\\
&\le C \|\varphi\|_\infty(c_0\ell(t-s))^{n-1} \ell(t-s)\le (c_1\ell(t-s))^n,\ \ n\ge 1,\end{split}\end{equation*} where $c_1:= c_0 + C\|\varphi\|_\infty.$ Since $\ell(r)\to 0$ as $r\to 0$, we may find   $t_0\in (0,1]$ such that $c_1\ell(t_0)<1$. Combining this with (\ref{FF1}) we conclude   from the construction of $p_{b,c}^{(\aa)}$ and the definition of $P_{t,s}^{b,c}$ that $\|\nn u(t,\cdot)\|_\infty$ is bounded on $[s,s+t_0].$ This and  (\ref{FF0}) yield
 \beq\label{WE3} \sup_{r\in [s,t_0+s]}\big\{\|\nn P_{r,s}^{b,c}\varphi\|_\infty+\|P_{r,s}^{b,c}\varphi\|_\infty\big\} <\infty.\end{equation}
Therefore,  there exists a constant $C>0$ such that
\begin{align*}
\|P_{t,s}^{b,c}\varphi-\varphi\|_\infty&\leq \|P_{t-s}^{(\aa)}\varphi -\varphi\|_\infty
+\bigg\|\int^t_s\!\!\!\int_M p^{(\aa)}(t-r,\cdot,z)|c(r,z)|\cdot|u(r,z)|\mu(\d z)\d r\bigg\|_\infty\\
&\qquad+\bigg\|\int^t_s\!\!\!\int_M p^{(\aa)}(t-r,\cdot,z)|b(r,z)|\cdot|\nabla_zu(r,z)|\mu(\d z)\d r\bigg\|_\infty\\
&\leq\|P_{t-s}^{(\aa)}\varphi -\varphi\|_\infty+C\sup_{r\in[s,t]}\|P_{r,s}^{b,c}\varphi\|_\infty K^{0,0}_{\alpha,c}(t-s)\\
&\qquad+C\sup_{r\in[s,t]}\|\nabla P_{r,s}^{b,c}\varphi\|_\infty K^{0,0}_{\alpha,|b|}(t-s)\to 0,
\end{align*}
as $t\downarrow s$. Hence,
\begin{align}
\lim_{t\downarrow s}\sup_{r\in[s,t]}\|P^{b,c}_{r,s}\varphi-\varphi\|_\infty
 =0.\label{WE4}
\end{align}
Since  $b\in C([0,\infty); L_{loc}^1(M\to TM;\mu)),c\in C([0,\infty); L_{loc}^1(M\to\R;\mu)),$ this and (\ref{WE3}) yield that
$$\lim_{t\downarrow s}  \int_M\psi \big\{I_1(t,s)+I_2(t,s)\big\}\d\mu =0.$$
Combining this with (\ref{WE}) and (\ref{WE2}), we need only to prove
\beq\label{WE5} \lim_{t\downarrow s} \int_M \psi I_3(t,s)\d\mu=0.\end{equation}
To this end, take $\{b_n(s,\cdot)\}_{n\ge 1}\subset C_0^\infty(M;TM)$ such that
$$\lim_{n\to\infty} \int_M |\psi| \cdot|b_n(s,\cdot)-b(s,\cdot)|\d\mu =0.$$
Since $\mu$ has a $C^1$-density w.r.t. the volume measure, $\div_\mu(\psi b_n(s))\in C_0(M)$ for $\psi\in C_0^2(M)$.
Combining this with (\ref{WE3}) and (\ref{WE4}), and by the dominated convergence theorem we conclude that
\begin{align*}
&\limsup_{t\downarrow s} \bigg|\int_M\psi I_3(t,s)\d\mu \bigg|\\
&\leq \limsup_{n\to\infty}\limsup_{t\downarrow s} \frac{1}{t-s}\int^t_s\!\!\!\int_M|b(s,\cdot)-b_n(s,\cdot)|\cdot|\nabla P^{b,c}_{r,s} \varphi-\nabla\varphi|\cdot|\psi|\d\mu \d r\\
&\quad+\limsup_{n\to\infty}\limsup_{t\downarrow s}\frac{1}{t-s}\int^t_s\!\!\!\int_M| \div_\mu (b_n(s,\cdot)\psi) |\cdot |P^{b,c}_{r,s} \varphi -\varphi |\d\mu\d r=0.
\end{align*}
Then the  proof is complete.
\end{proof}

\beg{proof}[Proof of Theorem \ref{1.2}] Due to Theorem \ref{T1.1}(1), it suffices to prove for $0\le s<t$ with $t-s\le t_0$, where $t_0>0$ is fixed in the proof of Theorem \ref{T1.1}.
According to (a)  in the proof of Theorem \ref{T1.1}, we only need to prove
\begin{align}\label{EU1} &\Big|\nabla_x  \Theta_n(t,x;s,y)-\parallel_{x'\to x} \nabla_{x'}\Theta_n(t,x';s,y)\Big| \\
&\quad\leq(c_0\ell_\beta(t-s))^n(t-s)^{-\frac{\beta}{\alpha}}\rr(x,x')^{\beta-1}\eta(t-s;x,x';y), n\geq 1,\nonumber
\end{align}
where
$$
\ell_\beta(r):=\sup_{\vv\in (0, r]} \big\{K_{\aa,|b|}^{\beta,\beta}(\vv)+K_{\aa,c}^{\beta,\beta}(\vv)\big\},\ \ r>0.
$$
By (\ref{1.4}), Lemma \ref{L3.1} and Proposition \ref{PP3}(2),(3), we have
\begin{align*}
&\Big|\nabla_x\Theta_n(t,x;s,y)-\parallel_{x'\to x}\nabla_{x'}\Theta_n(t,x';s,y)\Big|\\
& \leq\int^t_s\!\!\!\int_M\Big(|b(r,z)|\cdot|\nabla_z \Theta_n(r,z;s,y)|+|c(r,z)|\cdot|\Theta_n(r,z;s,y)|\Big)\\
&\qquad\qquad\times\Big|\nabla_x p^{(\aa)}(t-r,x,z)-\parallel_{x'\to x}\nabla_{x'} p^{(\aa)}(t-r,x',z)\Big|\mu(\d z)\d r\\
&\quad\leq C\{c_0\ell(t-s)\}^n \rr(x,x')^{\beta-1}\int^t_s\!\!\!\int_M\Big[\eta (t-r;x,x';z)p^{(\aa)}(r-s,z,y)\\
&\qquad\qquad\times\Big(|b(r,z)|(r-s)^{-\frac{1}{\alpha}}(t-r)^{-\frac{\beta}{\alpha}}
+|c(r,z)|(t-r)^{-\frac{\beta}{\alpha}}\Big)\Big]\mu(\d z)\d r.
\end{align*}
By the (3P)-inequality, we have
\begin{align*}
&\eta (t-r;x,x';z)p^{(\aa)}(r-s,z,y)\\
&\leq p^{(\aa)}(t-s,x,y)\Big\{p^{(\aa)}(t-r,x,z)+p^{(\aa)}(r-s,z,y)\Big\}\\
&\quad+p^{(\aa)}(t-s,x',y)\Big\{p^{(\aa)}(t-r,x',z)+p^{(\aa)}(r-s,z,y)\Big\}\\
&\quad+\int^1_0p^{(\aa)}(t-s,\gamma^{x,x'}_\theta,y)\Big\{p^{(\aa)}(t-r,\gamma^{x,x'}_\theta,z)+p^{(\aa)}(r-s,z,y)\Big\}\d\theta.
\end{align*}
Substituting this into the above estimate, we find that
\begin{align*}
&\Big|\nabla_x\Theta_n(t,x;s,y)-\parallel_{x'\to x}\nabla_{x'}\Theta_n(t,x';s,y)\Big|\\
&\leq C \{c_0\ell(t-s)\}^n\rr(x,x')^{\beta-1}
\eta (t-s;x,x';y)(t-s)^{-\frac{\beta}{\alpha}}\\
&\quad\times \Big(K^{1,\beta}_{\alpha,|b|}(t-s)+K^{\beta,1}_{\alpha,|b|}(t-s)
+K^{0,\beta}_{\alpha,c}(t-s)+K^{\beta,0}_{\alpha,c}(t-s)\Big).
\end{align*}This implies (\ref{EU1}). \end{proof}

\beg{proof}[Proof of Theorem \ref{1.3}] By (\ref{A2}) and the existence of $\div_\mu b,$
we see that $\div_\mu\{p^{(\aa)}(t-r, x,\cdot)b(r,\cdot)\}$ exists. Take $\{h_m\}_{m\ge 1}\subset C_0^\infty(M)$
with $0\le h_m\le 1, h_m\uparrow 1, \|\nn h_m\|_\infty\downarrow 0$.
Then, by approximating $\Theta_{n-1}$ with $h_m \Theta_{n-1}$, we obtain from (\ref{2.3}) that
\begin{align*}
\Theta_n(t,x;s,y)&=\int^t_s\!\!\!\int_M\div_\mu\{p^{(\alpha)}(t-r,x,\cdot) b(r,\cdot)\}(z)\Theta_{n-1}(r,z;s,y)\mu(\d z)\d r\\
&\quad+\int^t_s\!\!\!\int_Mp^{(\alpha)}(t-r,x,z) c(r,z)\Theta_{n-1}(r,z;s,y)\mu(\d z)\d r\\
&=\int^t_s\!\!\!\int_M\<b(r,z),\nabla_zp^{(\alpha)}(t-r,x,z)\>\Theta_{n-1}(r,z;s,y)\mu(\d z)\d r\\
&\quad+\int^t_s\!\!\!\int_Mp^{(\alpha)}(t-r,x,z) \tilde c(r,z)\Theta_{n-1}(r,z;s,y)\mu(\d z)\d r,
\end{align*}
where
$$
\tilde c(r,z)=c(r,z)+\div_\mu b(r,z).
$$
Notice that by the symmetry and (\ref{1.3}),
$$
|\nn_y p(t,x,y)|=|\nn_y p(t,y,x)|\le \ff{C\exp[-\ff{\rr(x,y)^2}{Ct}]}{\ss t\,(\mu(M)\land t)^{\ff d 2} },\ \ t>0, x,y\in M.
$$
Using the same arguments as in Lemma \ref{L3.1}, one can prove
\begin{align*}
|\nabla_y \Theta_n(t,x;s,y)| \leq \{c_0\tilde\ell(t-s)\}^n(t-s)^{-\frac{1}{\alpha}}p^{(\aa)}(t-s,x,y),
\end{align*}
where
$$
\tilde\ell(r):=\sup_{\vv\in (0, r]} \big\{K_{\aa,|b|}^{1,1}(\vv)+K_{\aa,\tilde c}^{1,1}(\vv)\big\},\ \ r>0.
$$
As in the proof of Theorem \ref{T1.1}, $\sum_{n=0}^\infty \nn_y  \Theta_n(t,x;s,y)$
converges  uniformly on $\{(t,x;s,y):\ t-s\in (0,t_0], x,y\in M\}.$ Thus (\ref{A22}) holds.
\end{proof}

\section{Some examples}

\paragraph{Example 4.1.} Let   $L=\DD+ \<\nn V, \nn \cdot\>$
for some $V\in C^2(M)$ such that
\beg{equation}\label{4.1} \beg{split} &\Ric(X,X)-\Hess_V(X,X) -\vv \<X,\nn V\>^2\ge 0,\ \ X\in TM,\\
 &\ff{r^d}{C}\le \mu(B(x,r))\le C r^d,\ \ r>0, x\in M\end{split}\end{equation}
 hold  for some constants $\vv>0, C>1,$ where $B(x,r)$ is the geodesic ball at $x$ with radius $r$, and $\mu(\d x)=\e^{V(x)}{\rm vol}(\d x) $ for ${\rm vol}$   the volume measure on $M$. Then $P_t$ is symmetric in $L^2(\mu)$ and all assertions in  Theorems \ref{T1.1} hold.
In fact, according to \cite{Q} (see also \cite{LY} when $V=0$), the condition (\ref{1.2}) follows from (\ref{4.1}).
Next,  according to e.g. \cite[Corollary 4.2]{W97}
\beq\label{4.2}|\nn P_t f|\le \ff {C(P_t f^2)^{\ff 1 2} }{\ss t},\ \ t>0, f\in \B_b(M)\end{equation} holds for some constant $C>0$. Letting $f_{t,y}(z)= p(\ff t 2, z,y)$, this  implies that
$$|\nn_x p(t,x,y)|=|\nn P_{\frac{t}{2}}f_{t,y}(x)|
\le \ff{C(P_{\ff t 2} f^2_{t,y})^{\ff 1 2}} {\ss t}\le \ff{C\ss{\|f_{t,y}\|_\infty p(t,x,y)}}{\ss{1\land t}}.$$ Combining this with (\ref{1.2}) we prove   (\ref{1.3}).

\paragraph{Example 4.2.} Let $M$ be compact and $L=\DD+\<\nn V, \nn \cdot\>$ for some $V\in C^2(M)$. Let $\mu(\d x)=\e^{V(x)}{\rm vol}(\d x).$  Then all assertions in Theorem  \ref{T1.2} hold.
In this case   $\ff{(1\land r)^d}{C}\le \mu(B(x,r))\le C (r\land 1)^d$ holds for some constants $C>1$ and all $x\in M, r>0.$ So, (\ref{1.2}) follows from   \cite{LY} or \cite{Q}.
Next, since the compactness of $M$ implies that $\Ric-\Hess_V$ is bounded below,    \cite[Corollary 4.2]{W97} implies (\ref{4.2}) for $t\in (0,1].$ Thus, as observed in Example 4.1 that (\ref{1.3}) holds for $t\in (0,1].$ Again since $M$ is compact, the second assertion in \cite[Theorem 4.4]{W97} implies that
$$|\nn_x p(t,x,y)|\le C\e^{-\ll t},\ \ t\ge 1, x,y\in M$$ holds for some constants $C,\ll>0.$ Therefore, (\ref{1.3}) holds also for $t\ge 1$ as $\rr$ is bounded.

\paragraph{Example 4.3.} Let $M=\R^d, \mu(\d x)=\d x$ and
$$L=\sum_{i,j=1}^d a_{ij}(x)\ff{\pp^2}{\pp x_i\pp x_j}.$$
Assume that  $a_{ij}$   are bounded and H\"older continuous functions on $\R^d$,  and $(a_{ij})\ge
\ll_0 I_{d\times d}$ holds for some constant $\ll_0>0.$  Then all assertions in Theorems \ref{T1.1} and \ref{T1.2} hold. In fact, (\ref{1.2}) follows from \cite[Theorem A]{Sheu} with $b=0$, and (\ref{1.3}) and (\ref{HESS}) follow from \cite[(1.3)]{Sheu} (see also \cite{FR}, page 229).

\paragraph{Acknowledgement.} The authors would like to thank Professor Renming Song for valuable conversations.

\end{document}